\documentclass[a4paper,11pt]{article}
\usepackage{jheppub}
\usepackage{array}
\usepackage{amssymb}
\usepackage{amsmath}
\usepackage{amsfonts}
\usepackage{mathrsfs}
\usepackage{graphics}
\usepackage{graphicx}
\usepackage{bm}
\usepackage{bbm}
\usepackage{dsfont}
\usepackage{tikz,ifthen}

\makeatletter
\def\@fpheader{\relax}
\makeatother

 \usetikzlibrary{trees}
 \usetikzlibrary{decorations.pathmorphing}
 \usetikzlibrary{decorations.markings}
 \tikzset{beamerprimary/.style={structure.fg, thick}}
 \tikzset{beamersecondary/.style={structure.bg, thick}}
 \tikzset{boson/.style={draw=structure.fg,decorate, decoration={snake}},
     gauge/.style={decorate, decoration={snake} },
     fermion/.style={postaction={decorate},
         decoration={markings,mark=at position .55 with {\arrow{>}}}},
     fermionloop/.style={postaction={decorate},
         decoration={markings,mark=at position .25 with {\arrow{<}}}},
     gluon/.style={decorate,
         decoration={coil,amplitude=1pt, segment length=2pt}},
     scalar/.style={dashed},
     graviton/.style={double},
     ghostloop/.style={postaction={decorate},
     decoration={densely dotted,markings,mark=at position .25 with {\arrow{<}}}},
     ghost/.style={densely dotted,postaction={decorate},
         decoration={markings,mark=at position .55 with {\arrow{>}}}}
 }

\newcommand{\bpsi}{\bar{\psi}}
\newcommand{\bq}{\bar{q}}

\newcommand{\tr}{\mathrm{tr}}

\newcommand{\ft}{\text{finite}}
\newcommand{\MS}{\overline{\text{MS}}}

\title{Renormalization of the second-order QCD with arbitrary chromomagnetic factor and quark self-interactions}
\author{Carlos A. Vaquera-Araujo}
\affiliation{Facultad de Ciencias, CUICBAS, Universidad de Colima,\\ Bernal D\'iaz del Castillo 340, Colima, M\'exico, 28045.}

\emailAdd{carolusvaquera@gmail.com}

\abstract{In this work, the renormalization of the second-order QCD is analyzed at one-loop level, including an arbitrary 
chromomagnetic factor and quark self-interactions, which are mass dimension four operators in this framework. 
The model is shown to be renormalizable for any value of the chromomagnetic factor $\kappa$ in the validity region of the perturbative expansion, and Dirac QCD is
recovered in the $\kappa=2$ case with vanishing self-interaction couplings.}

\begin{document}
\maketitle
\flushbottom

\section{Introduction}

The idea of a second order formalism for spin $1/2$ fermions can be traced back to Feynman \cite{feynman1} and Gell-Mann \cite{fg}. Applied to QCD, this formalism has proved to be very useful in the efficient determination of multi-gluon amplitudes due to its resemblance with scalar calculations (see for example \cite{schubert} and further references). Important developments can be found in \cite{veltman} and specially in \cite{morgan} for the non-abelian gauge theory. More recently, an analysis of the anomalous chromomagnetic moment couplings of the top quark was performed in \cite{Larkoski:2010am}, where a second order formalism was used.

In general, an anomalous magnetic moment can be implemented naturally in a second order theory, and besides, it can be properly renormalized at one-loop level in the abelian version if point-like four-fermion interactions  are included \cite{AN,VNA}. Notice that this is possible because fermion fields have mass dimension one in this case. The purpose of this work is to generalize this result to a non-abelian gauge theory, analyzing the one-loop renormalization of the QCD of second order fermions with an arbitrary chromomagnetic factor in presence of fermion self-interactions. The starting point of this paper is the Poincar\'e-projector formalism \cite{NKR,DNR}, a framework based on the projection onto irreducible representations of the Poincar\'e group, constructed to solve the problems of the quantum description of interacting high spin fields. 

Second order fermions are conceptually different to Dirac fermions. If no constraint is imposed, a second order fermion contains 8 dynamical degrees of freedom. As shown in \cite{veltman,morgan}, there is a consistent reduction of dynamical degrees of freedom and a direct connection between second order fermions and the Dirac formalism only if the chromomagnetic factor (or gyromagnetic in the abelian version) is set to the fixed value $\kappa=2$. Therefore, the results obtained in this work do not directly apply to Dirac fermions, but only to second order ones.  It is also unclear if the renormalizable theory described here corresponds to a perturbation theory about a sensible zeroth-order Hamiltonian. Further work must be done in order to elucidate the physical status of the model. However, in its present form, this theory might be useful as a suitable starting point for the formulation of effective field theories.

This paper is organized as follows: In Section \ref{sectMod}, the basics of the model are presented.  The renormalization procedure is sketched in section \ref{sectRen}. Section \ref{secDiv} is devoted to the calculation of the divergent piece of all the potentially divergent amplitudes of the theory and section \ref{bfss} to the determination of the corresponding beta functions. Finally, the conclusions of the work are discussed in section \ref{sectSumm}.

\section{The model}\label{sectMod}
\subsection{Lagrangian}\label{ssectLG}
The basic dimension four $SU(N_c)$ gauge invariant Lagrangian in the Poincar\'e-projector formalism is given by
\begin{equation}
\begin{split}
\mathscr{L}_{SU(N_c)}=& (\overline{D_\mu q})K^{\mu\nu}(D_\nu q)-m^2\bar{q}q  -\frac{1}{4}G^{a\mu\nu}G^a_{\mu\nu}, 
\label{NKRlag1}
\end{split}
\end{equation}
where $q$ are the quark fields, living in the
fundamental representation of $SU(N_c)$, with an internal $SU(N_f)$ global flavor symmetry (all flavors are degenerate with mass $m$) and covariant derivatives
\begin{equation}
D_\mu q= \left(\partial_\mu-ig A^a_\mu t^a\right)q.
\end{equation}
$A^a_\mu$ stands for the gluon field, $t^a$ are the generators of $SU(N_c)$ ($1\leq a\leq N_c^2-1$), and the field strength tensor is
\begin{equation}
G^{a}_{\mu\nu}=\partial_\mu A^a_\nu-\partial_\nu A^a_\mu+gf^{abc}A^b_{\mu}A^c_{\nu},
\end{equation}
with structure constants $f^{abc}$ defined by
\begin{equation}
[t^a,t^b]=if^{abc}.
\end{equation}
The space-time tensor $K^{\mu\nu}$ in the quark kinetic term is defined as
\begin{equation}
K^{\mu\nu}\equiv g^{\mu\nu}- i\kappa M^{\mu\nu} -i \kappa^{\prime}\tilde{M}^{\mu\nu},
\end{equation}
with the Lorentz generators $M^{\mu\nu}$ for the $(1/2,0)\oplus(0,1/2)$ representation, their dual generators $\tilde{M}^{\mu\nu}=\epsilon^{\mu\nu\alpha\beta}M_{\alpha\beta}/2$ and arbitrary constants $\kappa$, $\kappa'$.   Here $\kappa$ can be identified
with the chromomagnetic factor, and $\kappa^{\prime}$ parameterizes parity violating chromoelectric dipole interactions (see \cite{AN,VNA} for further details and conventions). Setting $\kappa=2$ and $\kappa^{\prime}=0$, eq.
(\ref{NKRlag1}) describes the second order formalism for QCD \cite{morgan}. In this paper, the analysis is restricted to $\kappa'=0$, with $\kappa$ kept arbitrary.

In order to obtain the gluon propagator from  eq. (\ref{NKRlag1}), the gauge is fixed through a covariant gauge contribution and the Faddeev-Popov ghosts $\bar{c}^a$, $c^a$ using
\begin{equation}
\mathscr{L}_{\text{gauge}}=-\frac{1}{2\xi}\left(\partial^\mu A^a_\mu\right)^2+\left(\partial^\mu\bar{c}^a\right)\left(D_\mu c^{a}\right),
\end{equation}
with the ghost covariant derivative: $D_\mu  c^{a}= \left(\partial_\mu\delta^{ac}+g f^{abc}A^{b}_\mu \right)c^{c} $, where the unit matrix in the adjoint representation $\delta^{ab}$ is written down explicitly.  Introducing an auxiliary scalar field $B^a$, this Lagrangian can be written, up to a total divergence, as
\begin{equation}
\mathscr{L}_{\text{gauge}}=-\frac{\xi}{2}(B^a)^2+B^a\partial^\mu A^a_\mu-\bar{c}^a\left(\partial^\mu D_\mu c^{a}\right),
\end{equation}

As stated in the introduction, second order fermion fields have mass dimension $1$ in four space-time dimensions, and therefore, also    Nambu-Jona-Lasinio-like interactions \cite{Nambu:1961tp,Nambu:1961fr}
are dimension-four and gauge-invariant. For a large number of quark flavors, the possible self-interactions (involving only elements of the $M^{\mu\nu}$ algebra) are:

\begin{equation}
\begin{split}
\mathscr{L}_{\text{self}}&=\frac{\lambda_{S}}{2}\left(\bar{q}q\right)^{2}+ \frac{\lambda_{P}}{2}\left(\bar{q}\gamma^5 q\right)^{2}+\frac{\lambda_{T}}{2}\left(\bar{q}M^{\mu\nu}q\right)^{2}\\
&+\frac{\lambda_{S_t}}{2}\left(\bar{q} t^a q\right)^{2}+ \frac{\lambda_{P_t}}{2}\left(\bar{q}\gamma^5t^a  q\right)^{2}+\frac{\lambda_{T_t}}{2}\left(\bar{q}M^{\mu\nu}t^a q\right)^{2}.
\label{NKRlag2}
\end{split}
\end{equation}

Thus, the complete Lagrangian of the model is
\begin{equation}
\begin{split}
\mathscr{L}=\mathscr{L}_{SU(N_c)}+\mathscr{L}_{\text{gauge}}+\mathscr{L}_{\text{self}}.
\label{NKRlag}
\end{split}
\end{equation}
Notice that this Lagrangian is BRST invariant, as can be checked with the transformations
\begin{equation}
\begin{split}
\delta A^{a}_{\mu}&=\theta D_{\mu}c^a,\\
\delta q&= ig\theta c^at^aq,\\
\delta c^a&=-\frac{1}{2}g\theta f^{abc}c^bc^c,\\
\delta \bar{c}^a&=\theta B^a,\\
\delta B^a&= 0, 
\label{BRST}
\end{split}
\end{equation}
parameterized by the infinitesimal anticommuting constant $\theta$.

\subsection{Feynman rules}\label{ssectFR}

\begin{figure}[ht]
\centering
\includegraphics[width=0.6\textwidth]{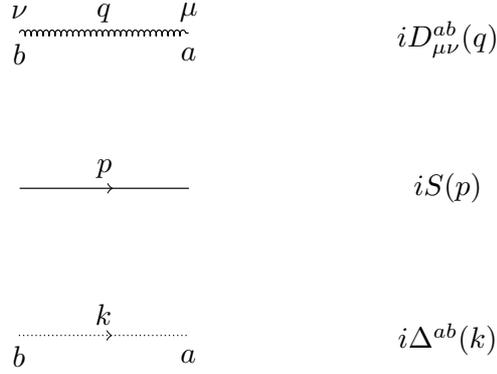}
\caption{Propagators.}
\label{FD1}
\end{figure}

The quadratic part of $\mathscr{L}$ gives the propagators shown in figure \ref{FD1}. Defining $\square\left[p\right]\equiv p^2-m^2+i\varepsilon$ and $\triangle\left[q\right]\equiv q^2	+i\varepsilon$, the quark propagator is 
\begin{equation}
iS(p) =\frac{i}{\square\left[p\right]} \mathds{1}_s\mathds{1}_c\mathds{1}_f,
\end{equation}
with spinor unit matrix $\mathds{1}_s$, color unit matrix $\mathds{1}_c$ and flavor unit matrix $\mathds{1}_f$.
The gluon propagator is
\begin{equation}
iD^{ab}_{\mu\nu}(q) =
-\frac{i}{\triangle\left[q\right]} \left[g_{\mu \nu }+\left(\xi-1\right)\frac{q_\mu q_\nu}{q^2}\right]\delta^{ab},
\end{equation}
and the ghost propagator is
\begin{equation}
iG^{ab}(k) =
\frac{i}{\triangle\left[k\right] }\delta^{ab}.
\end{equation}

\begin{figure}[ht]
\centering
\includegraphics[width=\textwidth]{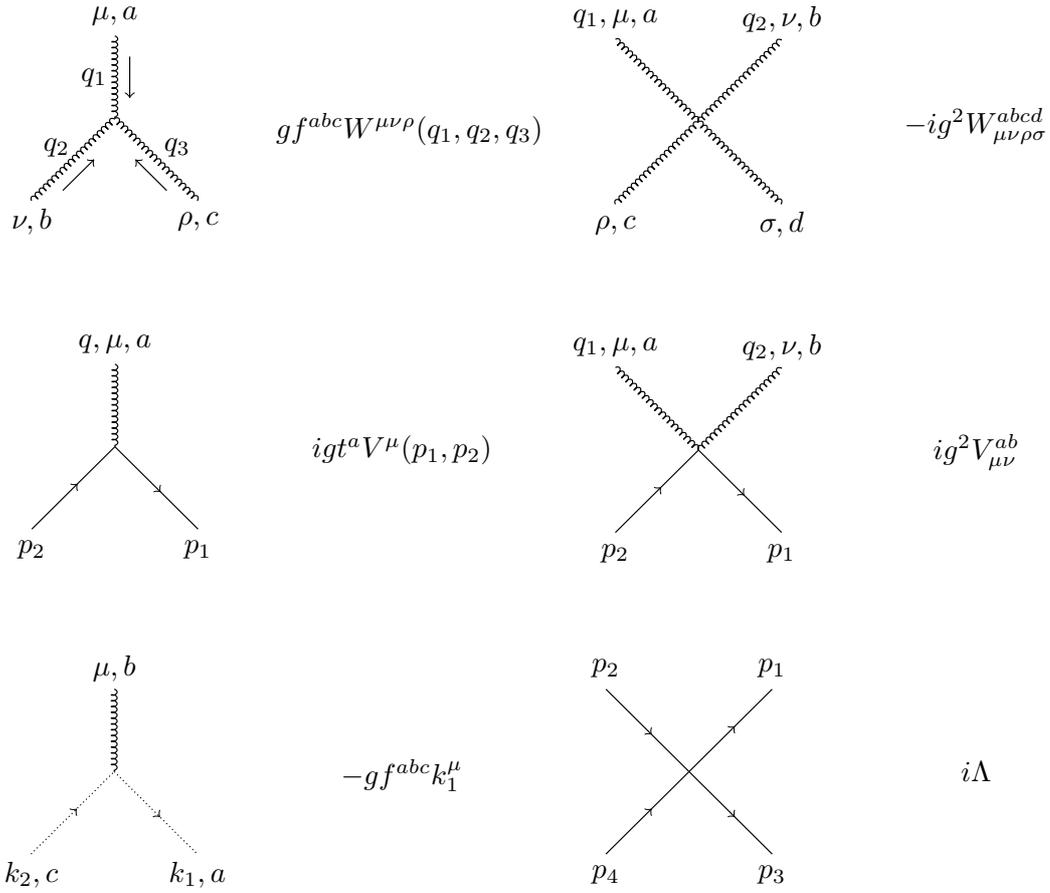}
\caption{Feynman rules for the second order QCD of self-interacting quarks with arbitrary chromomagnetic factor.}
\label{FD2}
\end{figure}

The Feynman rules for the interaction vertices of $\mathscr{L}$ are shown in figure \ref{FD2},
with the following definitions:
\begin{align}
W^{\mu\nu\rho}(q_1,q_2,q_3)=&(q_1-q_2)^\rho g^{\mu\nu}+(q_2-q_3)^\mu g^{\nu\rho}+(q_3-q_1)^\nu g^{\rho\mu},\\
W^{abcd}_{\mu\nu\rho\sigma}=&f^{abe}f^{cde}\left(g_{\mu\rho}g_{\nu\sigma}-g_{\mu\sigma}g_{\nu\rho}\right)+f^{ace}f^{bde}\left(g_{\mu\nu}g_{\rho\sigma}-g_{\mu\sigma}g_{\nu\rho}\right)\\&+f^{ade}f^{bce}\left(g_{\mu\nu}g_{\rho\sigma}-g_{\mu\rho}g_{\nu\sigma}\right),\nonumber\\
V^{\mu }(p^{\prime},p)=&\left[(p^{\prime}+p)^{\mu }\mathds{1}_s+i\kappa M^{\mu \nu }(p^{\prime}-p)_{\nu}\right]\mathds{1}_f,\\
V^{ab}_{\mu\nu }=&\left[ g_{\mu \nu }\{t^a,t^b\}\mathds{1}_s-i\kappa M_{\mu \nu }[t^a,t^b]\right]\mathds{1}_f.
\end{align}
The 4-quark vertex is given by
\begin{equation}
\begin{split}
i\Lambda&=i\left(\lambda_{S}\mathds{1}\overline{\otimes}\mathds{1}+\lambda_{P}\chi\overline{\otimes}\chi+\lambda_{T}M\overline{\otimes}M
+\lambda_{S_t}t\overline{\otimes}t+\lambda_{P_t}\chi_t\overline{\otimes}\chi_t+\lambda_{T_t}M_t\overline{\otimes}M_t\right),
\end{split}
\end{equation} 
where the following notation is adopted
\begin{equation}
\begin{split}
\mathds{1}\overline{\otimes}\mathds{1}=&\left[
\delta_{i_1i_2}\delta_{i_3i_4}\delta_{j_1j_2}\delta_{j_3j_4}\delta_{\alpha_1\alpha_2}\delta_{\alpha_3\alpha_4}
-\delta_{i_1i_4}\delta_{i_3i_2}\delta_{j_1j_4}\delta_{j_3j_2}\delta_{\alpha_1\alpha_4}\delta_{\alpha_3\alpha^{}_2}\right],\\
\chi\overline{\otimes}\chi=&\left[
\delta_{i_1i_2}\delta_{i_3i_4}\delta_{j_1j_2}\delta_{j_3j_4}\gamma^5_{\alpha_1\alpha_2}\gamma^5_{\alpha_3\alpha_4}
-\delta_{i_1i_4}\delta_{i_3i_2}\delta_{j_1j_4}\delta_{j_3j_2}\gamma^5_{\alpha_1\alpha_4}\gamma^5_{\alpha_3\alpha^{}_2}\right],\\
M\overline{\otimes}M=&\left[
\delta_{i_1i_2}\delta_{i_3i_4}\delta_{j_1j_2}\delta_{j_3j_4}M^{\mu\nu}_{\alpha_1\alpha_2}M_{\mu\nu\alpha_3\alpha_4}
-\delta_{i_1i_4}\delta_{i_3i_2}\delta_{j_1j_4}\delta_{j_3j_2}M^{\mu\nu}_{\alpha_1\alpha_4}M_{\mu\nu\alpha_3\alpha^{}_2}\right],\\
t\overline{\otimes}t=&\left[
\delta_{i_1i_2}\delta_{i_3i_4}t^a_{j_1j_2}t^a_{j_3j_4}\delta_{\alpha_1\alpha_2}\delta_{\alpha_3\alpha_4}
-\delta_{i_1i_4}\delta_{i_3i_2}t^a_{j_1j_4}t^a_{j_3j_2}\delta_{\alpha_1\alpha_4}\delta_{\alpha_3\alpha^{}_2}\right],\\
\chi_t\overline{\otimes}\chi_t=&\left[
\delta_{i_1i_2}\delta_{i_3i_4}t^a_{j_1j_2}t^a_{j_3j_4}\gamma^5_{\alpha_1\alpha_2}\gamma^5_{\alpha_3\alpha_4}
-\delta_{i_1i_4}\delta_{i_3i_2}t^a_{j_1j_4}t^a_{j_3j_2}\gamma^5_{\alpha_1\alpha_4}\gamma^5_{\alpha_3\alpha^{}_2}\right],\\
M_t\overline{\otimes}M_t=&\left[
\delta_{i_1i_2}\delta_{i_3i_4}t^a_{j_1j_2}t^a_{j_3j_4}M^{\mu\nu}_{\alpha_1\alpha_2}M_{\mu\nu\alpha_3\alpha_4}
-\delta_{i_1i_4}\delta_{i_3i_2}t^a_{j_1j_4}t^a_{j_3j_2}M^{\mu\nu}_{\alpha_1\alpha_4}M_{\mu\nu\alpha_3\alpha^{}_2}\right],
\end{split}
\end{equation}
for explicit flavor ($1\leq i_k \leq N_f$), color ($1\leq j_k \leq N_c$) and spinor ($1\leq \alpha_k \leq 4$) indices of the quarks labeled with
momenta $p_k$ ($1\leq k\leq 4$) in the corresponding diagram of figure \ref{FD2}. 
For closed quark loops, a factor of $-1$ given by Fermi
statistics must be included, as well as the appropriate symmetry factors for closed gluon loops. In second order QCD with $\kappa=2$, an additional $1/2$ factor can be attached
to each spinor space trace in order to provide a direct connection between the second order formalism and the Dirac one  \cite{veltman, morgan}. In this work the convention used in \cite{VNA}  is adopted, where the trace of the spinor unit matrix is written as $\tr[\mathds{1}_s]=\tau$, with $\tau=4$ for the second order formalism in its natural form, and the formal relation $\tau=2$ for comparison with Dirac QCD.   

The $SU(N_c)$ conventions used in this work are
\begin{equation}
\begin{array}{rclcrcl}
\tr[t^a t^b]&=&T_F\delta^{ab},&\qquad\qquad\qquad &&&\\
t^a t^a&=&C_F,& \qquad\qquad\qquad &C_F&=&T_F\left(N_c-\frac{1}{N_c}\right),\\
if^{acd}if^{bdc}&=&C_A\delta^{ab},&\qquad\qquad\qquad& C_A&=&2T_FN_c,
\end{array}
\end{equation} 
with $T_F$ as the normalization constant for generator traces, and $C_F$ ($C_A$) as the Casimir of the fundamental (adjoint) representation.

\section{Renormalization}\label{sectRen}

\subsection{Counterterms} \label{countertermsss}
The free and interacting pieces of the bare Lagrangian are
\begin{align}
\mathscr{L}_{\text{free}}=&-\frac{1}{4}\left(\partial^\mu A_0^{a\nu}-\partial^\nu A_0^{a\mu}\right)\left(\partial_\mu A^a_{0\nu}-\partial_\nu A^a_{0\mu}\right)- \frac{1}{2\xi_0}(\partial^\mu A_{0\mu})^2\\
&+\left(\partial^\mu\bar{c}_0^a\right)\left(\partial_\mu c_0^{a}\right)+ \left(\partial^\mu \bq_0 \right)\left(\partial_{\mu} q_0\right) -m_{0}^{2} \bq_0 q_0,\nonumber\\
\mathscr{L}_{\text{int}}=& -\frac{g}{2}f^{abc}\left(\partial^\mu A_0^{a\nu}-\partial^\nu A_0^{a\mu}\right) A_{0\mu}^{b}A_{0\nu}^{c}
\\&-\frac{g^2}{4}f^{eab}f^{ecd}A_{0\mu}^{a}A_{0\nu}^{b}A_{0}^{c\mu}A_{0}^{d\nu}-gf^{abc}\left(\partial_\mu\bar{c}_0^a\right)c_0^{b}A^{c\mu}\nonumber\\
&+ig_0 \left[ \bq_0   K_{0\nu\mu} \left(\partial^\mu q_0\right) - \left(\partial^\mu
\bpsi_0\right) K_{0\mu\nu}\psi_0   \right] A^{a \nu}_0t^a + g^{2}_0 \bq_0 K_{0\mu\nu} q_0 A^{a\mu}_0t^a A^{b\nu }_0t^b \nonumber\\
&+\frac{\lambda_{S0}}{2}\left(\bq_0q_0\right)^{2}+ \frac{\lambda_{P0}}{2}\left(\bq_0\gamma^5 q_0\right)^{2}+\frac{\lambda_{T0}}{2}\left(\bq_0M^{\mu\nu}q_0\right)^{2}\nonumber\\
&+\frac{\lambda_{S_t0}}{2}\left(\bq_0 t^a q_0\right)^{2}+ \frac{\lambda_{P_t0}}{2}\left(\bar{q_0}\gamma^5t^a  q_0\right)^{2}+\frac{\lambda_{T_t0}}{2}\left(\bq_0M^{\mu\nu}t^a q_0\right)^{2}.\nonumber
\end{align}
with $K^{\mu\nu}_{0}\equiv  g^{\mu\nu}-i\kappa_0 M^{\mu\nu}$.

Introducing the renormalized fields
\begin{equation}
A^{\mu}_r =  Z_A^{-\frac{1}{2}} A^{\mu}_0, \qquad  q_r =  Z_q^{-\frac{1}{2}} q_0, \qquad c_r =  Z_c^{-\frac{1}{2}} c_0,
\end{equation}
together with the renormalized parameters
\begin{equation}
\begin{array}{rclcrclcrcl}
g_r &=&  Z_g^{-1} g_0, &\qquad\qquad&  m_r &=&  Z_m^{-\frac{1}{2}} m_0, &\qquad\qquad& \kappa_r &=&  Z_\kappa^{-1} \kappa_0,\\
\lambda_{Sr} &=&  Z_S^{-1} \lambda_{S0}, &\qquad\qquad& \lambda_{Pr} &=&  Z_P^{-1} \lambda_{P0}, &\qquad\qquad& \lambda_{Tr} &=&  Z_T^{-1} \lambda_{T0},\\
\lambda_{S_tr} &=&  Z_{S_t}^{-1} \lambda_{S_t0}, &\qquad\qquad& \lambda_{P_tr} &=&  Z_{P_t}^{-1} \lambda_{P_t0}, &\qquad\qquad& \lambda_{T_tr} &=&  Z_{T_t}^{-1} \lambda_{T_t0},\\
&&&&\xi_r&=& Z_A^{-1}\xi_0,&&&&
\end{array}
\end{equation}
and the notation $Z\equiv 1+\delta$, the free  and interacting Lagrangians can be written as
\begin{align}
\mathscr{L}_{\text{free}}=&\mathscr{L}_{\text{free}}^{r}-\frac{1}{4}\left(\partial^\mu A_r^{a\nu}-\partial^\nu A_r^{a\mu}\right)\left(\partial_\mu A^a_{r\nu}-\partial_\nu A^a_{r\mu}\right) \delta_{A} \label{freeren}\\
&+\left(\partial^\mu\bar{c}_r^a\right)\left(\partial_\mu c_r^{a}\right)+\left(\partial^\mu\bar{c}_r^a\right)\left(\partial_\mu c_r^{a}\right)\delta_c\nonumber\\
&+\left(\partial^\mu \bq_r\right) \left(\partial_{\mu} q_r\right) \delta_{q}
-  m^2_r  \bq_r q_r \delta_{mq^2}  ,\nonumber\\
\mathscr{L}_{\text{int}}  =& \mathscr{L}^{r}_{\text{int}} -\frac{g_r}{2}f^{abc}\left(\partial^\mu A_r^{a\nu}-\partial^\nu A_r^{a\mu}\right) A_{r\mu}^{b}A_{r\nu}^{c}\delta_{A^3}\label{intren}\\
&-\frac{g_r^2}{4}f^{eab}f^{ecd}A_{r\mu}^{a}A_{r\nu}^{b}A_{r}^{c\mu}A_{r\nu}^{d\nu}\delta_{A^4}-g_rf^{abc}\left(\partial_\mu\bar{c}_r^a\right)c_r^{b}A^{c\mu}\delta_{Ac^2}\nonumber\\
& +ig_r \left[ \bq_r   g_{\nu\mu} \left(\partial^\mu q_r\right) - \left(\partial^\mu
\bq_r\right) g_{\mu\nu}q_r   \right] A^{a\nu}_rt^a \delta_{Aq^2 }\nonumber\\
&+ig_r \left[ \bq_r  (-i\kappa_r M_{\mu\nu})\left(\partial^\mu q_r\right) - \left(\partial^\mu
\bq_r\right)(-i\kappa_r M_{\mu\nu})q_r   \right] A^{a\nu}_rt^a \delta_{\kappa Aq^2}\nonumber\\
& + g^{2}_r \bq_r g_{\mu\nu} q_r A^{a\mu}_r t^a A^{b\nu}_rt^b \delta_{A^2q^2} +g^{2}_r \bq_r (-i\kappa_r M_{\mu\nu}) q_r A^{a\mu}_r t^a A^{b\nu}_rt^b \delta_{\kappa A^2q^2}\nonumber \\
&\frac{\lambda_{Sr}}{2}\delta_{Sq^4}\left(\bq_rq_r\right)^{2}+ \frac{\lambda_{Pr}}{2}\delta_{Pq^4}\left(\bq_r\gamma^5 q_r\right)^{2}+\frac{\lambda_{Tr}}{2}\delta_{Tq^4}\left(\bq_rM^{\mu\nu}q_r\right)^{2}\nonumber\\
&+\frac{\lambda_{S_tr}}{2}\delta_{S_tq^4}\left(\bq_r t^a q_r\right)^{2}+ \frac{\lambda_{P_tr}}{2}\delta_{P_tq^4}\left(\bar{q_r}\gamma^5t^a  q_r\right)^{2}+\frac{\lambda_{T_tr}}{2}\delta_{T_tq^4}\left(\bq_rM^{\mu\nu}t^a q_r\right)^{2},\nonumber
\end{align}
where $\mathscr{L}^{r}_{\text{free}}$ ($\mathscr{L}^{r}_{\text{int}}$) is the same Lagrangian as $\mathscr{L}_{\text{free}}$ ($\mathscr{L}_{\text{int}}$) with renormalized fields and couplings. The new renormalization constants introduced in eqs. (\ref{freeren}) and (\ref{intren}) are defined as follows:
\begin{equation}
\begin{array}{rclcrclcrcl}
Z_{mq^2}&=& Z_{m}Z_{q},&\qquad\qquad& Z_{A^3}&=& Z_g Z_A^{\frac{3}{2}}, &\qquad\qquad&
Z_{A^4}&=&  Z_g^2 Z_A^2, \\
Z_{Ac^2}&=&Z_gZ_A^{\frac{1}{2}}Z_c ,&\qquad\qquad& Z_{Aq^2}&=&Z_g Z_A^{\frac{1}{2}}Z_q , &\qquad\qquad&
Z_{A^2q^2}&=& Z^2_g Z_A Z_q, \\
Z_{\kappa Aq^2}&=&Z_{\kappa} Z_g Z_A^{\frac{1}{2}}Z_q, &\qquad\qquad&
Z_{\kappa A^2q^2}&=&Z_{\kappa}  Z^2_g Z_A Z_q ,&\qquad\qquad&
Z_{Sq^4}&=&Z_{S}Z_{q}^{2},\\
Z_{Pq^4}&=&Z_{P}Z_{q}^{2},&\qquad\qquad&
Z_{Tq^4}&=&Z_{T}Z_{q}^{2},&\qquad\qquad&
Z_{S_tq^4}&=&Z_{S_t}Z_{q}^{2},\\
Z_{P_tq^4}&=&Z_{P_t}Z_{q}^{2},&\qquad\qquad&
Z_{T_tq^4}&=&Z_{T_t}Z_{q}^{2}.&\qquad\qquad&
\end {array}
\label{count1}
\end{equation}

\begin{figure}[ht]
\centering
\includegraphics[width=0.6\textwidth]{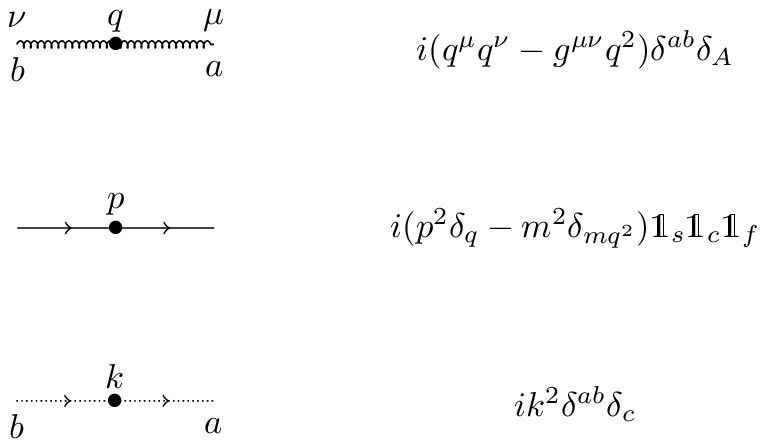}
\caption{Counterterms for Propagators.}
\label{CT1}
\end{figure}

\begin{figure}[ht]
\centering
\includegraphics[width=\textwidth]{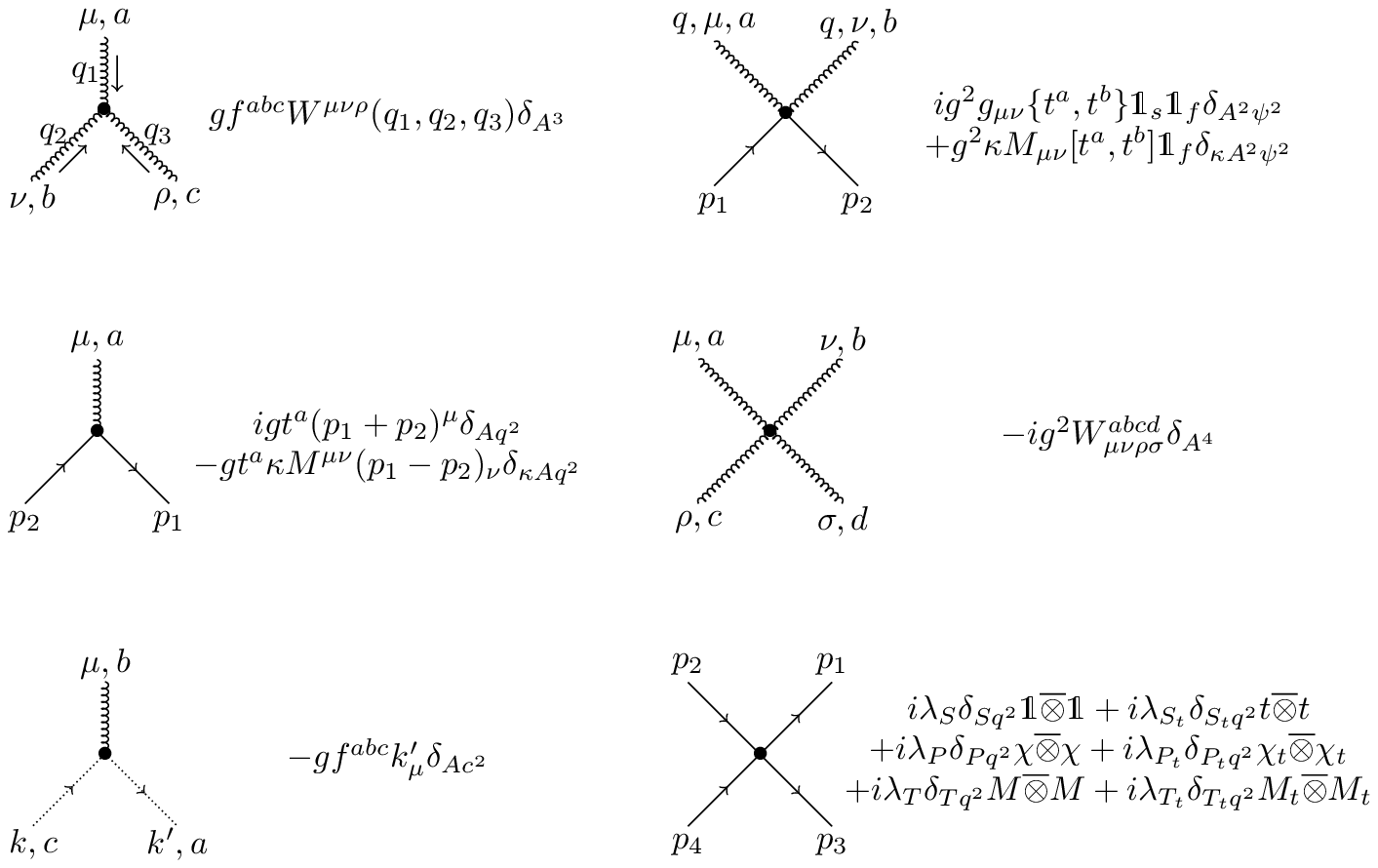}
\caption{Counterterms for 3-point and 4-point vertices.}
\label{CT2}
\end{figure}

In the following, for simplicity, the suffix $r$ will be removed in the renormalized parameters, keeping the
suffix $0$ for the bare quantities. In this notation, the Feynman rules for the renormalized fields are given in figures \ref{FD1} and \ref{FD2}, while the corresponding rules for the counterterms are shown in figures \ref{CT1} and \ref{CT2}.

\subsection{Superficial degree of divergence}
The superficial degree of divergence $D$ for this theory is analogous to that of scalar QCD (SQCD). A Feynman diagram with
$N_G$ external gluon lines, $N_F$ external quark lines and $N_{FP}$ ghost lines has
\begin{equation}
D\leq 4- N_G-N_F-\frac{3}{2}N_{FP}.
\end{equation} 
The factor $3/2$ takes into account the fact that half of the ghost internal lines adjacent to ghost external lines in the Feynman diagram do not contribute to 
$D$ due to the peculiar form of the ghost-ghost-gluon vertex.   

The eight superficially divergent Feynman amplitudes are the gluon, ghost and quark self-energy parts and the 3-gluon, ghost-ghost-gluon, quark-quark-gluon, 4-gluon and quark-quark-gluon-gluon vertices. The renormalization of these vertices will be carried at one-loop level
for arbitrary covariant gauge in dimensional regularization and in the $\overline{\text MS}$ scheme, using a commuting $\gamma^5$ with $M^{\mu\nu}$ in $d=4-2\epsilon$ dimensions.
As in \cite{VNA}, $\gamma^5$ is present only in diagrams with self-interactions, and therefore,
pure QCD diagrams are free from dimensional regularization inconsistencies. The same observations made in \cite{VNA} regarding self-interactions are applicable in this work, and further study must be done in this direction.  

In $d$ dimensions, the renormalized parameters scale according to
\begin{equation}
\begin{array}{rclcrclcrcl}
g_r&\rightarrow & \mu^\epsilon g_r,  &\qquad\qquad&
m_r&\rightarrow & m_r,&\qquad\qquad&
\kappa_r&\rightarrow & \kappa_r, \\
\lambda_{Sr}&\rightarrow & \mu^{2\epsilon} \lambda_{Sr}, &\qquad\qquad&
\lambda_{Pr}&\rightarrow & \mu^{2\epsilon} \lambda_{Pr}, &\qquad\qquad&
\lambda_{Tr}&\rightarrow & \mu^{2\epsilon} \lambda_{Tr},\\
\lambda_{S_tr}&\rightarrow & \mu^{2\epsilon} \lambda_{S_tr}, &\qquad\qquad&
\lambda_{P_tr}&\rightarrow & \mu^{2\epsilon} \lambda_{P_tr}, &\qquad\qquad&
\lambda_{T_tr}&\rightarrow & \mu^{2\epsilon} \lambda_{T_tr}.\\
\end {array}
\label{ddim}
\end{equation}

In the next section, the divergent pieces of the superficially divergent amplitudes are extracted in order to determine the value of the renormalization constants at one-loop precision.

\section{Divergent amplitudes}\label{secDiv}
\subsection{Gluon self-energy}

\begin{figure}
\centering
\includegraphics[width=0.8\textwidth]{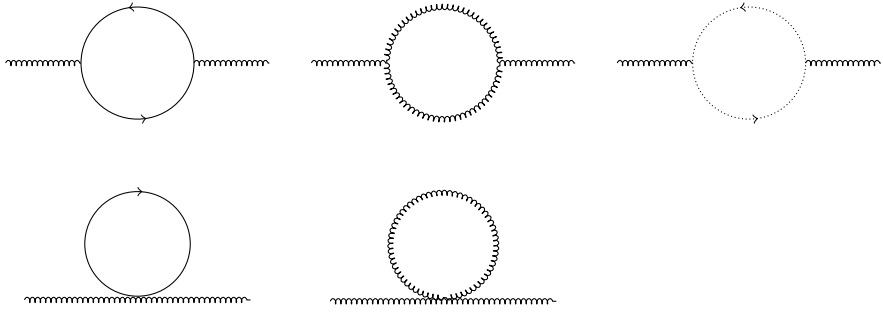}
\caption{Feynman diagrams for the gluon self-energy at one-loop.}
\label{ggfig}
\end{figure}

The gluon self-energy at one-loop level is
\begin{equation}
-i\Pi^{ab}_{\mu\nu}(q)=-i\Pi^{*ab}_{\mu\nu}(q)-i\delta_{A} \left(g_{\mu\nu}q^2-q_{\mu}q_{\nu}\right)\delta^{ab},
\label{vacuo}
\end{equation}
where $-i\Pi^{*ab}_{\mu\nu}(q)$ stands for the contribution of the one-loop diagrams in Figure \ref{ggfig}.
As expected, $\Pi^{ab}_{\mu\nu}(q)$ is transverse
\begin{equation}
\Pi^{ab}_{\mu\nu}(q)= (g^{\mu\nu}q^2-q^\mu q^\nu)\delta^{ab} \Pi(q^2) ,
\end{equation}
and the divergent structure of the $\Pi(q^2)$ form factor is given by
\begin{equation}
\Pi(q^2)=\frac{  g^2} {(4\pi)^2}\left[\tau N_f T_F \left(\frac{3 \kappa^2-4}{12}\right)-C_A \left(\frac{13}{6}
- \frac{\xi}{2}\right)\right]\frac{1}{\tilde{\epsilon}}+\delta_{A}+\text{finite}.
\end{equation}
Thus, in the $\overline{\text{MS}}$ subtraction scheme, the gluon-field renormalization constant at one loop is
\begin{equation}\label{ZA}
Z_{A}=1-\frac{  g^2} {(4\pi)^2}\left[\tau N_f T_F \left(\frac{3 \kappa^2-4}{12}\right)-C_A \left(\frac{13}{6}
- \frac{\xi}{2}\right)\right]\frac{1}{\tilde{\epsilon}}.
\end{equation}

\subsection{Fermion self-energy}
\begin{figure}[t]
\begin{center}
\includegraphics[width=0.7\textwidth]{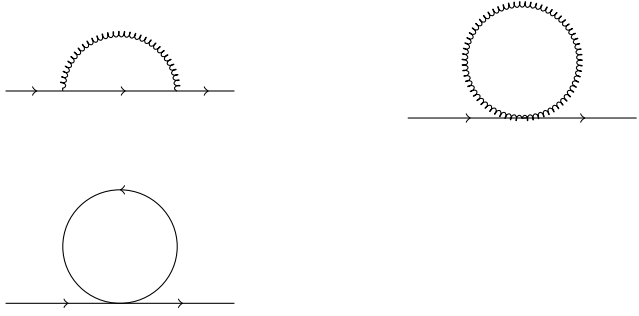}
\end{center}
\caption{Feynman diagrams for the fermion self-energy at one-loop.}
\label{self}
\end{figure}

The fermion self-energy at one-loop level reads
\begin{equation}
-i\Sigma(p^2)= -i \Sigma^{*}(p^2) + ip^2\delta_{q}- im^2 \delta_{mq^2},  \label{renoself}
\end{equation}
where $-i\Sigma^{*}(p^2)$ is computed with the one-loop diagrams of figure \ref{self} and yields
\begin{equation}
\begin{split}
\Sigma(p^2)=p^2\Sigma_1(p^2)+m^2\Sigma_2(p^2).
\end{split}
\end{equation}
where the form factors are
\begin{align}
\Sigma_1(p^2)=&\frac{g^2C_F(3-\xi)}{(4\pi)^2} \frac{1}{\tilde\epsilon}-\delta_{q}+ \ft,\\
\Sigma_2(p^2)=&\frac{1}{(4\pi)^2}\bigg\{g^2 C_F\left(\xi+ \frac{3\kappa^{2}}{4}\right)+C_F\left[\lambda_{S_t}+\lambda_{P_t}+3\lambda_{T_t}\right]\\&\qquad\quad- \left[
(\tau N_fN_C-1)\lambda_{S}-\lambda_{P}-3\lambda_{T}
\right]\bigg\} \frac{1}{\tilde\epsilon}+ \delta_{mq^2}+\ft.\nonumber
\end{align}
Form these results, the $Z_q$ and $Z_{mq^2}$ renormalization constants in the $\MS$ scheme are
\begin{align}
Z_q=&1+\frac{g^2C_F(3-\xi)}{(4\pi)^2} \frac{1}{\tilde\epsilon},\label{Zq}\\
Z_{mq^2}=&1-\frac{1}{(4\pi)^2}\bigg\{g^2 C_F\left(\xi+ \frac{3\kappa^{2}}{4}\right)+C_F\left[\lambda_{S_t}+\lambda_{P_t}+3\lambda_{T_t}\right]\label{Zmq^2}\\&\qquad\qquad\quad- \left[
(\tau N_fN_C-1)\lambda_{S}-\lambda_{P}-3\lambda_{T}
\right]\bigg\} \frac{1}{\tilde\epsilon}.\nonumber
\end{align}

It is important to point out that the quark mass is in general quadratically ultraviolet divergent in this model, a feature inherited from the close resemblance of this formalism to SQCD.

\subsection{Ghost self-energy}
\begin{figure}[t]
\begin{center}
\includegraphics[width=0.3\textwidth]{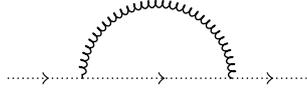}
\end{center}
\caption{Feynman diagram for the ghost self-energy at one-loop.}
\label{ghostself}
\end{figure}

The ghost self-energy at one-loop level is
\begin{equation}
-i\widetilde{\Sigma}^{ab}(k^2)= -i \widetilde{\Sigma}^{*ab}(k^2) + ik^2\delta^{ab} \delta_{c},
\end{equation}
where $\widetilde{\Sigma}^{*ab}(k^2)$ corresponds to the diagram of figure \ref{ghostself}. A direct calculation gives

\begin{align} 
\widetilde{\Sigma}^{ab}(p^2)=& \delta^{ab}k^2\left[\frac{g^2 C_A }{(4\pi)^2}\left(\frac{3-\xi}{4}\right)\frac{1}{\tilde\epsilon}-\delta_c\right]+\ft,
\end{align}
and the ghost-field $\MS$ renormalization constant is determined as
\begin{align} \label{Zc}
Z_c=& 1+\frac{g^2 C_A }{(4\pi)^2}\left(\frac{3-\xi}{4}\right)\frac{1}{\tilde\epsilon}.
\end{align}

\subsection{Quark-quark-gluon vertex}
\begin{figure}[t]
\begin{center}
\includegraphics[width=.7\textwidth]{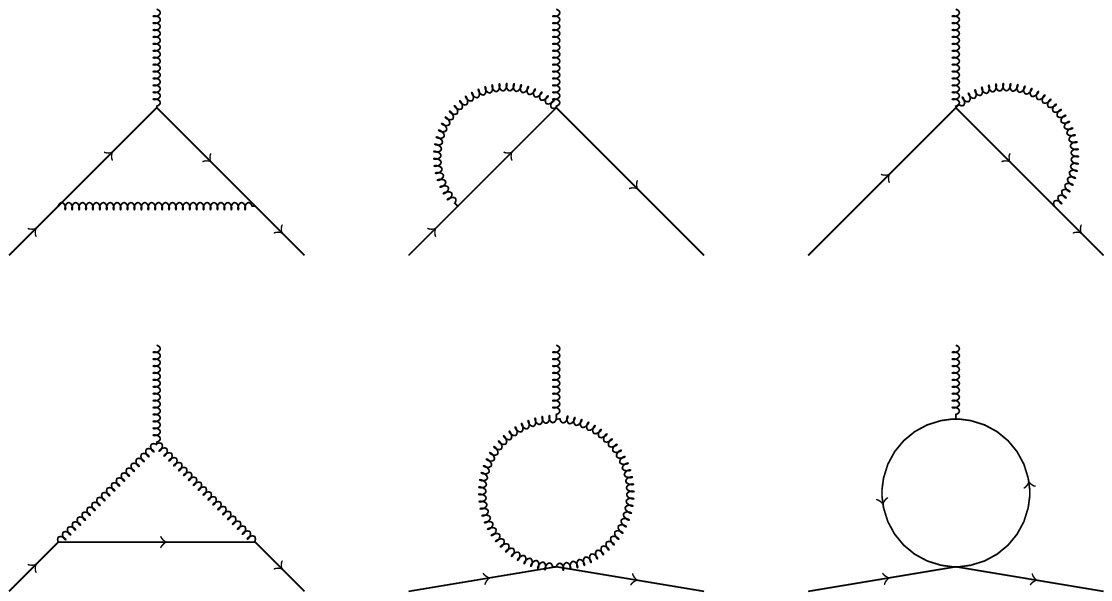}
\end{center}
\caption{Feynman diagrams for the $qqg$ vertex at one-loop.}
\label{ver3lazo}
\end{figure}
From the Feynman rules, the $qqg$ vertex at one-loop level is
\begin{align}
ig \Gamma^{a\mu}(p_1,p_2,q)=&igt^a V^{\mu}(p_1,p_2) +ig\Gamma^{*a\mu}(p_1,p_2,q)
\\&+igt^a V^{\mu}(p_1,p_2)\delta_{Aq^2}  +igt^a[ i\kappa M^{\mu\nu}(p_1-p_2)_\nu] \delta_{\kappa Aq^2},\nonumber
\end{align}
where $ig\Gamma^{*a\mu}(p',p,q)$ stands for the contribution of the one-loop diagrams in figure \ref{ver3lazo}. The divergent piece of this amplitude lives in two form factors
\begin{align}
\Gamma^{a\mu}(p_1,p_2,q)=t^a (p_1+p_2)^{\mu}\Gamma_1 +it^a\kappa M^{\mu\nu}(p_1-p_2)_\nu \Gamma_2 +\ft,
\end{align}
given by
\begin{align}
\Gamma_1=&-\frac{ g^2 }{(4 \pi) ^2} \left[C_F(3-\xi)-C_A\left(\frac{3+\xi}{4}\right)\right] \frac{1}{\tilde{\epsilon}}+\delta_{Aq^2}+\ft,\\
\Gamma_2=&\frac{ 1}{(4 \pi) ^2}\bigg\{\frac{ g^2}{4}\left[\frac{C_A}{2} \left(\kappa^2+4\kappa-6+2\xi\right)-C_F\left(\kappa^2+8-4\xi\right)\right]+\lambda_{S}+\lambda_{P}-\lambda_{T}\\&\qquad\quad+\left(C_F-\frac{C_A}{2}\right)\left[\lambda_{S_t}+\lambda_{P_t}-\lambda_{T_t}\right] +\frac{\tau}{2}N_fT_F\lambda_{T_t}\bigg\}\frac{1}{\tilde{\epsilon}}+\delta_{\kappa Aq^2}+\ft.\nonumber
\end{align}
Therefore, the $Z_{Aq^2}$ and $Z_{\kappa Aq^2}$ renormalization constants in the $\MS$ scheme are
\begin{align}
Z_{Aq^2}=&1+\frac{ g^2 }{(4 \pi) ^2} \left[C_F(3-\xi)-C_A\left(\frac{3+\xi}{4}\right)\right] \frac{1}{\tilde{\epsilon}},\label{ZAq^2}\\
Z_{\kappa Aq^2}=&1-\frac{ 1}{(4 \pi) ^2}\bigg\{\frac{ g^2}{4}\left[\frac{C_A}{2} \left(\kappa^2+4\kappa-6+2\xi\right)-C_F\left(\kappa^2+8-4\xi\right)\right]\label{ZkAq^2}
\\&\qquad\qquad\quad+\lambda_{S}+\lambda_{P}-\lambda_{T} +\frac{\tau}{2}N_fT_F\lambda_{T_t} \nonumber\\
&\qquad\qquad\quad+\left(C_F-\frac{C_A}{2}\right)\left[\lambda_{S_t}+\lambda_{P_t}-\lambda_{T_t}\right]\bigg\}\frac{1}{\tilde{\epsilon}}.\nonumber
\end{align}

\subsection{Ghost-ghost-gluon vertex}
\begin{figure}[t]
\begin{center}
\includegraphics[width=.5\textwidth]{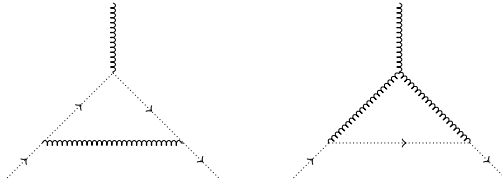}
\end{center}
\caption{Feynman diagrams for the $ccg$ vertex at one-loop.}
\label{ver3gh}
\end{figure}

The $ccg$ vertex at one-loop level is unmodified by the second order formalism for fermions
\begin{align}
-g\widetilde{\Gamma}^{abc\mu}(k_1,k_2,q)=-gf^{abc} k_1^{\mu} -g\widetilde{\Gamma}^{*abc\mu}(k_1,k_2,q)
-gf^{abc} k_1^{\mu}\delta_{ Ac^2},
\end{align}
with $-g\widetilde{\Gamma}^{*abc\mu}(k',k,q)$ calculated from the one-loop diagrams in figure \ref{ver3gh}. The divergent part of this amplitude is
\begin{equation}
\begin{split}
\widetilde{\Gamma}^{abc\mu}(k',k,q)=&f^{abc} k_1^{\mu}\left[\frac{g^2}{(4 \pi)^2}\frac{C_A}{2}\xi\frac{1}{\tilde{\epsilon}}+\delta_{ Ac^2}\right]+\ft,
\end{split}
\end{equation}
and the $Z_{Ac^2}$ $\MS$ renormalization constant reads
\begin{equation}\label{ZAc^2}
\begin{split}
Z_{ Ac^2}=&1-\frac{g^2}{(4 \pi)^2}\frac{C_A}{2}\xi\frac{1}{\tilde{\epsilon}}.
\end{split}
\end{equation}

\subsection{Three-gluon vertex}
\begin{figure}[t]
\begin{center}
\includegraphics[width=.7\textwidth]{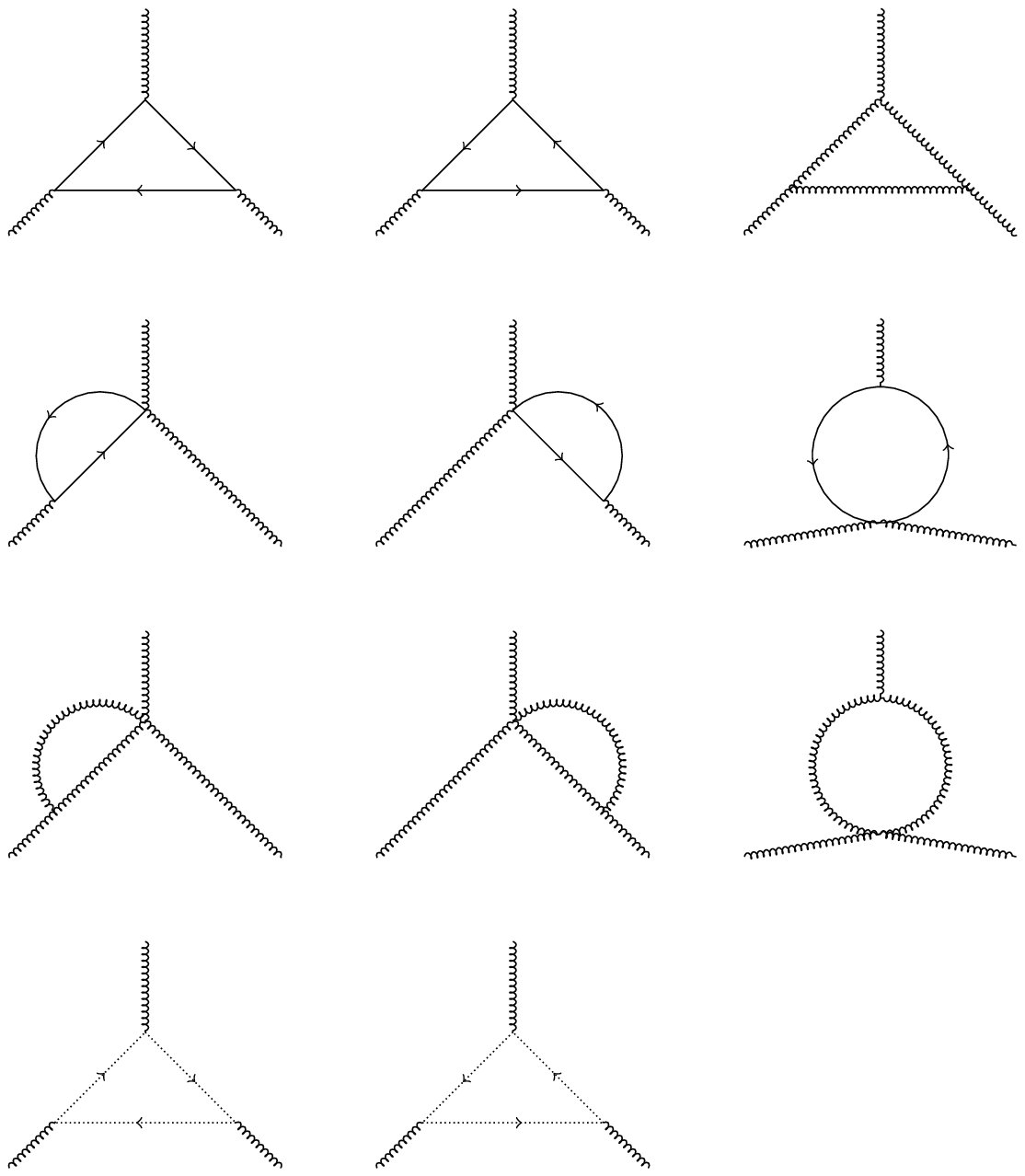}
\end{center}
\caption{Feynman diagrams for the $ggg$ vertex at one-loop.}
\label{gggfig}
\end{figure}
The $ggg$ vertex at one-loop level is
\begin{align}
g \mathcal{W}^{abc}_{\mu\nu\rho}(q_1,q_2,q_3)=&g f^{abc}W_{\mu\nu\rho}(q_1,q_2,q_3)+g \mathcal{W}^{*abc}_{\mu\nu\rho}(q_1,q_2,q_3)+g f^{abc}W_{\mu\nu\rho}(q_1,q_2,q_3)\delta_{A^3},
\end{align}
where $g \mathcal{W}^{*abc}_{\mu\nu\rho}(q_1,q_2,q_3)$ is the contribution of the one-loop diagrams in figure \ref{gggfig}. The divergent piece of this amplitude is contained in one form factor
\begin{align}
\mathcal{W}^{abc}_{\mu\nu\rho}(q_1,q_2,q_3)= f^{abc}W_{\mu\nu\rho}(q_1,q_2,q_3)\mathcal{W} +\ft,
\end{align}
given by
\begin{equation}
\begin{split}
\mathcal{W}=&\frac{ g^2 }{(4 \pi) ^2} \left[\tau N_f T_F \left(\frac{3 \kappa^2-4}{12}\right)-C_A \left(\frac{17}{12}
- \frac{3\xi}{4}\right)\right] \frac{1}{\tilde{\epsilon}}+\delta_{A^3}+\ft,
\end{split}
\end{equation}
which, in turn, determines the $Z_{A^3}$ renormalization constant in the $\MS$ scheme as
\begin{equation}\label{ZA^3}
\begin{split}
Z_{A^3}=&1-\frac{ g^2 }{(4 \pi) ^2} \left[\tau N_f T_F \left(\frac{3 \kappa^2-4}{12}\right)-C_A \left(\frac{17}{12}
- \frac{3\xi}{4}\right)\right] \frac{1}{\tilde{\epsilon}}.
\end{split}
\end{equation}

\subsection{Quark-quark-gluon-gluon vertex}
\begin{figure}[t]
\begin{center}
\includegraphics[width=\textwidth]{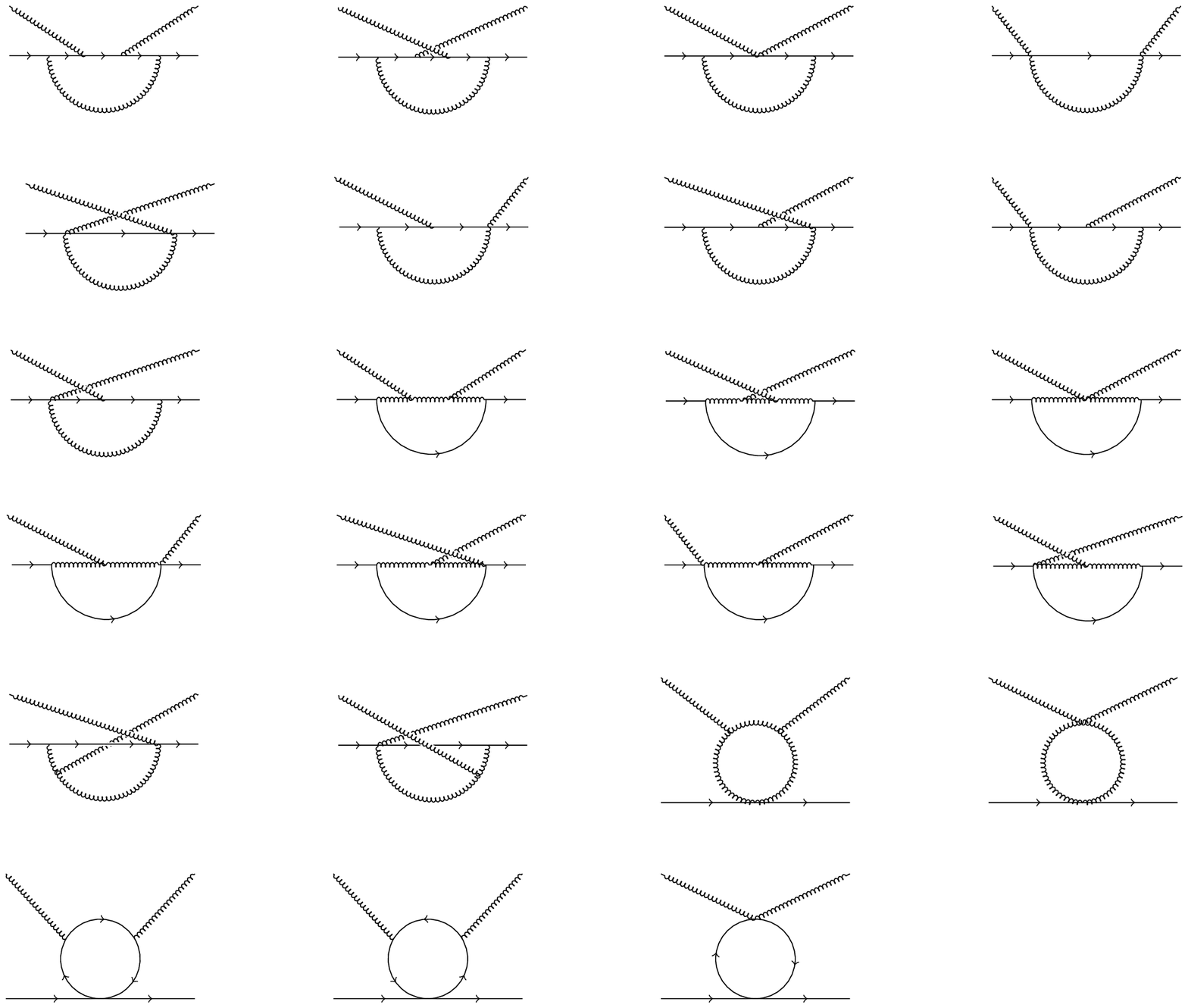}
\end{center}
\caption{Feynman diagrams for the $qqgg$ vertex at one-loop.}
\label{compton1}
\end{figure}
The $qqgg$ vertex function at one-loop level is obtained from the Feynman rules  as
\begin{align}
ig^2 \mathcal{G}^{ab}_{\mu\nu}(p_1,p_2,q_1,q_2) =& ig^2 V^{ab}_{\mu\nu}+ig^2 \mathcal{G}^{*ab}_{\mu\nu}(p_1,p_2,q_1,q_2) \\&+ ig^2 g_{\mu\nu}\{t^a,t^b\}\delta_{A^2q^2}-ig^2\left[i\kappa M_{\mu\nu}\right][t^a,t^b]\delta_{\kappa A^2q^2},\nonumber
\end{align}
where the one-loop corrections $ig^2 \mathcal{G}^{*ab}_{\mu\nu}(p_1,p_2,q_1,q_2)$ are given by the diagrams of figure \ref{compton1}. As expected, the divergent pieces of this amplitude are contained in two form factors
\begin{align}
\mathcal{G}^{ab}_{\mu\nu}(p_1,p_2,q_1,q_2) = g_{\mu\nu}\{t^a,t^b\}\mathcal{G}_1-i\kappa M_{\mu\nu}[t^a,t^b]\mathcal{G}_2+\ft,
\end{align}
read as
\begin{align}
\mathcal{G}_1=&-\frac{ g^2 }{(4 \pi) ^2} \left[C_F(3-\xi)-C_A\left(\frac{3+\xi}{2}\right)\right] \frac{1}{\tilde{\epsilon}}+\delta_{A^2q^2}+\ft,\\
\mathcal{G}_2=&\frac{ 1}{(4 \pi) ^2}\bigg\{\frac{ g^2}{4}\left[\frac{C_A}{2} \left(\kappa^2+4\kappa+4\xi\right)-C_F\left(\kappa^2+8-4\xi\right)\right]+\lambda_{S}+\lambda_{P}-\lambda_{T}\\&\qquad\quad+\left(C_F-\frac{C_A}{2}\right)\left[\lambda_{S_t}+\lambda_{P_t}-\lambda_{T_t}\right] +\frac{\tau}{2}N_fT_F\lambda_{T_t}\bigg\}\frac{1}{\tilde{\epsilon}}+\delta_{\kappa A^2q^2}+\ft.\nonumber
\end{align}
From these results, the $Z_{A^2q^2}$ and $Z_{\kappa A^2q^2}$ renormalization constants in the $\MS$ scheme are given by
\begin{align}
Z_{A^2q^2}=&1+\frac{ g^2 }{(4 \pi) ^2} \left[C_F(3-\xi)-C_A\left(\frac{3+\xi}{2}\right)\right] \frac{1}{\tilde{\epsilon}},\label{ZA^2q^2}\\
Z_{\kappa A^2q^2}=&1-\frac{ 1}{(4 \pi) ^2}\bigg\{\frac{ g^2}{4}\left[\frac{C_A}{2} \left(\kappa^2+4\kappa+4\xi\right)-C_F\left(\kappa^2+8-4\xi\right)\right]\label{ZkA^2q^2}\\
&\qquad\qquad\quad+\lambda_{S}+\lambda_{P}-\lambda_{T} +\frac{\tau}{2}N_fT_F\lambda_{T_t} \nonumber\\
&\qquad\qquad\quad+\left(C_F-\frac{C_A}{2}\right)\left[\lambda_{S_t}+\lambda_{P_t}-\lambda_{T_t}\right]\bigg\}\frac{1}{\tilde{\epsilon}}.\nonumber
\end{align}

\subsection{Four-gluon vertex}
\begin{figure}[t]
\begin{center}
\includegraphics[width=.8\textwidth]{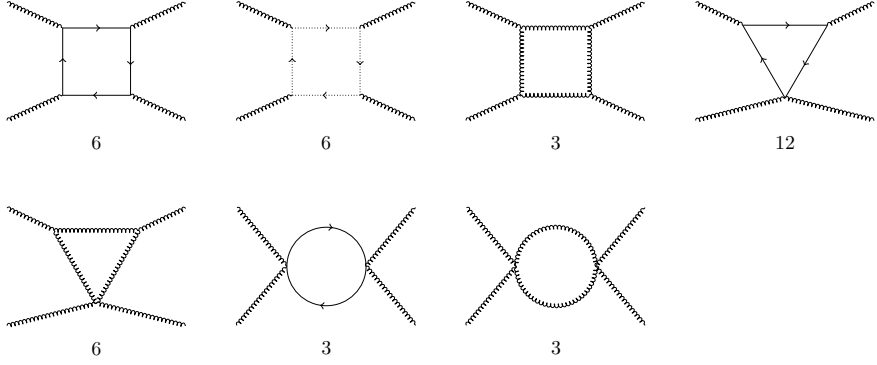}
\end{center}
\caption{Basic Feynman diagrams for the $gggg$ vertex at one-loop. The number indicates the total number of diagrams obtained from each basic form under permutation of external gluon lines.}
\label{ggggfig}
\end{figure}
The $gggg$ vertex at one-loop level is
\begin{align}
-ig^2 \mathcal{E}^{abcd}_{\mu\nu\rho\sigma}(q_1,q_2,q_3,q_4)=&-ig^2 W^{abcd}_{\mu\nu\rho\sigma}-ig^2 \mathcal{E}^{*abcd}_{\mu\nu\rho\sigma}(q_1,q_2,q_3,q_4)-ig^2 W^{abcd}_{\mu\nu\rho\sigma}\delta_{A^4},
\end{align}
where $-ig^2 \mathcal{E}^{abcd}_{\mu\nu\rho\sigma}(q_1,q_2,q_3,q_4)$ is the contribution of the one-loop diagrams in figure \ref{ggggfig}. The divergent piece of this amplitude is contained in one form factor
\begin{align}
\mathcal{E}^{abcd}_{\mu\nu\rho\sigma}(q_1,q_2,q_3,q_4)= W^{abcd}_{\mu\nu\rho\sigma}\mathcal{E}+\ft,
\end{align}
given by
\begin{equation}
\begin{split}
\mathcal{E}=&\frac{ g^2 }{(4 \pi) ^2} \left[\tau N_f T_F \left(\frac{3 \kappa^2-4}{12}\right)-C_A \left(\frac{2}{3}
-\xi\right)\right] \frac{1}{\tilde{\epsilon}}+\delta_{A^4}+\ft,	
\end{split}
\end{equation}
and the corresponding $\MS$ renormalization constant yields
\begin{equation}
\begin{split}\label{ZA^4}
Z_{A^4}=&1-\frac{ g^2 }{(4 \pi) ^2} \left[\tau N_f T_F \left(\frac{3 \kappa^2-4}{12}\right)-C_A \left(\frac{2}{3}
-\xi\right)\right] \frac{1}{\tilde{\epsilon}}.	
\end{split}
\end{equation}

\subsection{Four-quark vertex}
\begin{figure}[t]
\begin{center}
\includegraphics[width=\textwidth]{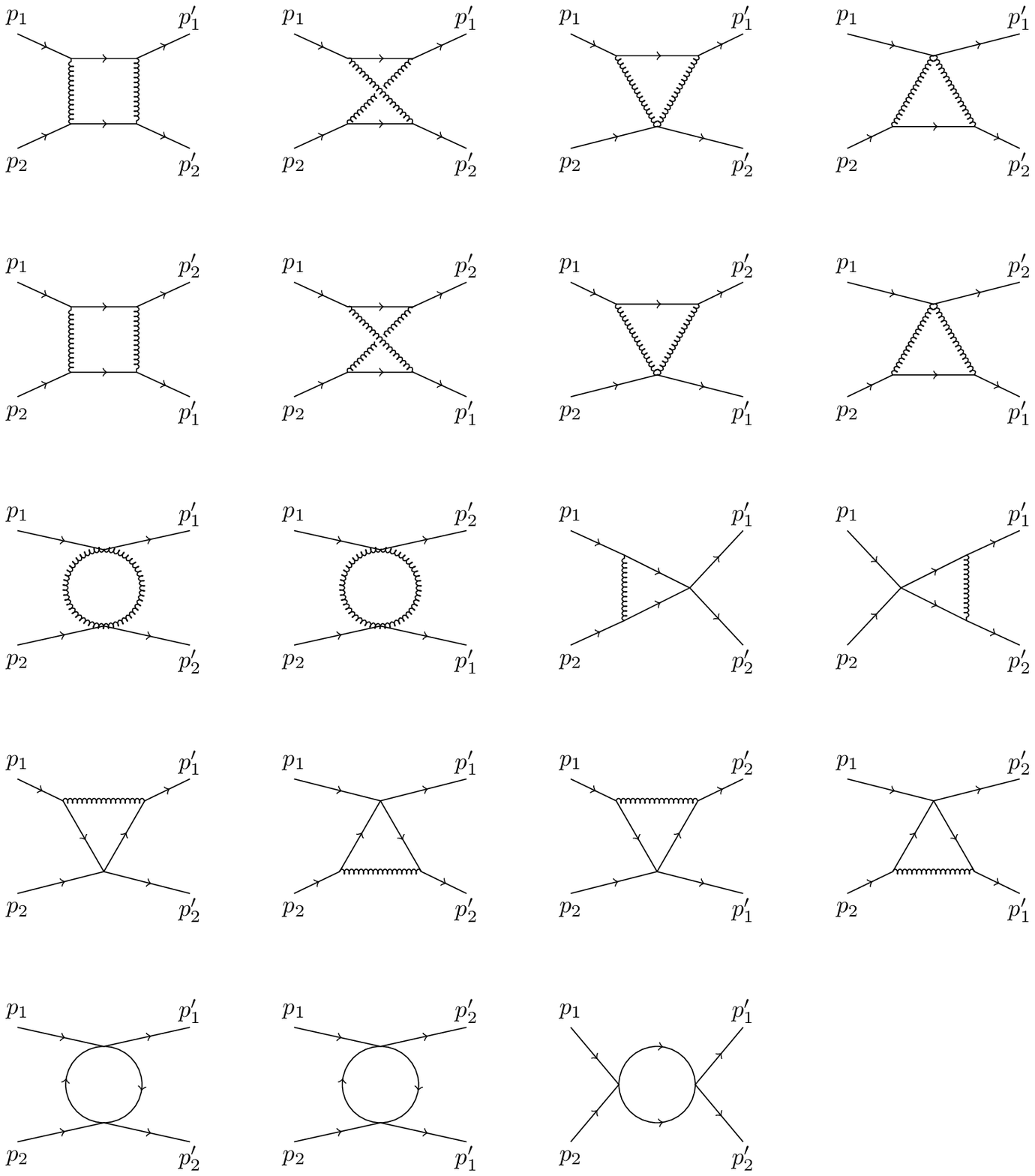}
\end{center}
\caption{Feynman diagrams for the $qqqq$ vertex function at one-loop.}
\label{ffff}
\end{figure}
The last potentially divergent vertex, is the
$qqqq$ vertex function 
\begin{equation}
\label{4fr}
\begin{split}
&i\mathcal{F}\left(p_1,p_2,p'_1,p'_2\right)=i\Lambda+i \mathcal{F}^{*}\left(p_1,p_2,p'_1,p'_2\right)+i\lambda_{S}\delta_{Sq^2}\mathds{1}\overline{\otimes}\mathds{1}+i\lambda_{S_t}\delta_{S_tq^2}t\overline{\otimes}t\\
&\qquad
+i\lambda_{P}\delta_{Pq^2}\chi\overline{\otimes}\chi+i\lambda_{P_t}\delta_{P_tq^2}\chi_t\overline{\otimes}\chi_t+i\lambda_{T}\delta_{Tq^2}M\overline{\otimes}M
+i\lambda_{T_t}\delta_{T_tq^2}M_t\overline{\otimes}M_t,
\end{split}
\end{equation}
with $i\mathcal{F}^{*}\left(p_1,p_2,p'_1,p'_2\right)$ obtained from the loop diagrams in figure
\ref{ffff}.  The divergent piece of this amplitude lives in six form factors
\begin{equation}
\begin{split}
\mathcal{F}\left(p_1,p_2,p'_1,p'_2\right)=&\mathcal{F}_1\mathds{1}\overline{\otimes}\mathds{1}+\mathcal{F}_2t\overline{\otimes}t
+\mathcal{F}_3\chi\overline{\otimes}\chi+\mathcal{F}_4\chi_t\overline{\otimes}\chi_t\\&+\mathcal{F}_5M\overline{\otimes}M
+\mathcal{F}_6M_t\overline{\otimes}M_t+\ft,
\end{split}
\end{equation}
given by
\begin{align}
\mathcal{F}_1=&\frac{1}{(4\pi)^2}\Bigg\{C_F\left(C_A-2C_F\right)\left[3g^4\left(1-\frac{\kappa^2}{4}\right)^2+\frac{3g^2\kappa^2}{4}\lambda_{T_t}+ \lambda_{S_t}^2+ \lambda_{P_t}^2+\frac{3}{2} \lambda_{T_t}^2\right]\\&
\qquad\quad+  2C_F \lambda_{S}\left(2	g^2\xi+\frac{3g^2\kappa^2}{4}+\lambda_{S_t}+\lambda_{P_t}+3
   \lambda_{T_t}\right)\nonumber
   \\&
\qquad\quad+  \lambda_{S} \left[ \lambda_{S}(4- \tau N_fN_c)+2\lambda_{P}+6 \lambda_{T}\right] +2 \lambda_{P}^2
   +3 \lambda_{T}^2
\Bigg\}\frac{1}{\tilde{\epsilon}}+\lambda_{S}\delta_{Sq^2}+\ft,\nonumber\\
\mathcal{F}_2=&\frac{1}{(4\pi)^2}\bigg\{\left(4C_F-\frac{3}{2}C_A\right)\left[3g^4\left(1-\frac{\kappa^2}{4}\right)^2+\frac{3g^2\kappa^2}{4}\lambda_{T_t}+\lambda_{S_t}^2+ \lambda_{P_t}^2+\frac{3}{2} \lambda_{T_t}^2\right]\\&
\qquad\quad+\lambda_{S_t}\left[2C_Fg^2\xi+\left(2C_F-C_A\right)\left(\frac{3g^2\kappa^2}{4}+\lambda_{S_t}+\lambda_{P_t} +3\lambda_{T_t}\right)\right]\nonumber
   \\&
\qquad\quad+\lambda_{S_t}\left[6 (\lambda_{S}+ \lambda_{T})+2\lambda_{P}
   -\tau N_f T_F \lambda_{S_t}\right]\nonumber\\&
\qquad\quad   +\lambda_{T}\left(\frac{3g^2\kappa^2}{2}+6\lambda_{T_t}\right)+
   4 \lambda_{P}\lambda_{P_t}
\bigg\}\frac{1}{\tilde{\epsilon}}+\lambda_{S_t}\delta_{S_tq^2}+\ft,\nonumber
\\
\mathcal{F}_3=&\frac{1}{(4\pi)^2}\bigg\{C_F\left(C_A-2C_F\right)\left[\frac{3g^2\kappa^2}{4}\lambda_{T_t}+2 \lambda_{P_t} \lambda_{S_t}+\frac{3}{2}
   \lambda_{T_t}^2\right]\\&
\qquad\quad+2C_F\lambda_{P}\left(2g^2\xi+\frac{3g^2\kappa^2}{4}+\lambda_{S_t}+\lambda_{P_t}+3 \lambda_{T_t}\right)\nonumber
\\&
\qquad\quad+\lambda_{P}\left[ \lambda_{P} (2-\tau N_f N_c )+6 (\lambda_{S}+\lambda_{T})\right]
+3 \lambda_{T}^2
 \bigg\}\frac{1}{\tilde{\epsilon}}+\lambda_{P}\delta_{Pq^2}+\ft,\nonumber
\end{align}
\begin{align} 
\mathcal{F}_4=&\frac{1}{(4\pi)^2}\bigg\{\left(4C_F-\frac{3}{2}C_A\right)\left[\frac{3g^2\kappa^2}{4}\lambda_{T_t}+2\lambda_{S_t}\lambda_{P_t}+\frac{3}{2} \lambda_{T_t}^2\right]\\&
\qquad\quad+\lambda_{P_t}\left[2C_Fg^2\xi+\left(2C_F-C_A\right)\left(\frac{3g^2\kappa^2}{4}+\lambda_{S_t}+\lambda_{P_t} +3\lambda_{T_t}\right)\right]
\nonumber   \\&
\qquad\quad+\lambda_{P_t}\left[6 (\lambda_{S}+\lambda_{T})+2\lambda_{P}
   -\tau N_f T_F \lambda_{P_t}\right]\nonumber\\&
\qquad\quad   +\lambda_{T}\left(\frac{3g^2\kappa^2}{2}+6\lambda_{T_t}\right)+
   4 \lambda_{P}\lambda_{S_t}
\bigg\}\frac{1}{\tilde{\epsilon}}+\lambda_{P_t}\delta_{P_tq^2}+\ft,\nonumber
\end{align}
\begin{align} 
\mathcal{F}_5=&\frac{1}{(4\pi)^2}\Bigg\{C_F\left(C_A-2C_F\right)\left(\frac{g^2\kappa^2}{2}+ 2\lambda_{T_t} \right)\left(\lambda_{S_t}+\lambda_{P_t}\right)\\&
\qquad\quad+  2C_F\lambda_{T} \left(2g^2\xi-\frac{g^2\kappa^2}{4}+\lambda_{P_t}
   +\lambda_{S_t}- \lambda_{T_t}\right)\nonumber
\\&
\qquad\quad+  \lambda_{T} \left[6 ( \lambda_{S}+ \lambda_{P})-2\lambda_{T}\left(1+
   \frac{\tau}{4}N_f N_c\right)\right]\Bigg\}\frac{1}{\tilde{\epsilon}}+\lambda_{T}\delta_{Tq^2}+\ft,\nonumber\\
\mathcal{F}_6=&\frac{1}{(4\pi)^2}\bigg\{C_Ag^2\left[\frac{g^2\kappa^2}{4}\left(1-\frac{\kappa}{2}\right)^2-\frac{3\kappa^2}{4}\left(\lambda_{S_t}+\lambda_{P_t}-\lambda_{T_t}\right)\right]+g^2\kappa^2\left(\lambda_{S}+\lambda_{P}\right)
\\&
\qquad\quad+C_Fg^2\left[2\lambda_{T_t}\xi+\frac{\kappa^2}{2}\left(4\lambda_{S_t}+4\lambda_{P_t}-\lambda_{T_t}\right)\right]+4 \lambda_{T}(\lambda_{S_t}+\lambda_{P_t} ) \nonumber\\&
\qquad\quad	+2C_A\lambda_{T_t}\left(\lambda_{T_t}-2\lambda_{S_t}-2\lambda_{P_t}\right)+2C_F\lambda_{T_t}\left(5\lambda_{S_t}+5\lambda_{P_t}-\lambda_{T_t}\right)	\nonumber\\
&\qquad\quad+\lambda_{T_t}\left[6\left(\lambda_{S}+\lambda_{P}\right)-2\lambda_{T}-\frac{\tau}{2}N_fT_F\lambda_{T_t}\right]
\bigg\}\frac{1}{\tilde{\epsilon}}+\lambda_{T_t}\delta_{T_tq^2}+\ft.\nonumber
\end{align}

Finally, the $\MS$ renormalization constants associated to this vertex are determined as
\begin{align}
Z_{Sq^2}=&1-\frac{1}{(4\pi)^2}\Bigg\{
 2C_F \left(2	g^2\xi+\frac{3g^2\kappa^2}{4}+\lambda_{S_t}+\lambda_{P_t}+3
   \lambda_{T_t}\right)\label{ZSq^2}
   \\&
\qquad\qquad\quad + C_F\left(C_A-2C_F\right)\frac{g^2}{\lambda_{S}}\left[3g^2\left(1-\frac{\kappa^2}{4}\right)^2+\frac{3\kappa^2}{4}\lambda_{T_t}\right]\nonumber\\&
\qquad\qquad\quad + C_F\left(C_A-2C_F\right)\frac{1}{\lambda_{S}}\left(\lambda_{S_t}^2+ \lambda_{P_t}^2+\frac{3}{2} \lambda_{T_t}^2\right)	\nonumber\\&
\qquad\qquad\quad+   \lambda_{S}(4- \tau N_fN_c)+2\lambda_{P}+6 \lambda_{T} +2 \frac{\lambda_{P}^2}{\lambda_{S}}
   +3 \frac{\lambda_{T}^2}{\lambda_{S}}
\Bigg\}\frac{1}{\tilde{\epsilon}},\nonumber
\\
Z_{S_tq^2}=&1-\frac{1}{(4\pi)^2}\bigg\{
2C_Fg^2\xi+\left(2C_F-C_A\right)\left(\frac{3g^2\kappa^2}{4}+\lambda_{S_t}+\lambda_{P_t} +3\lambda_{T_t}\right)\label{ZStq^2}
   \\&\qquad\qquad\quad +	\left(4C_F-\frac{3}{2}C_A\right)\frac{g^2}{\lambda_{S_t}}\left[3g^2\left(1-\frac{\kappa^2}{4}\right)^2+\frac{3\kappa^2}{4}\lambda_{T_t}\right]\nonumber
 \\&\qquad\qquad\quad +	\left(4C_F-\frac{3}{2}C_A\right)\frac{1}{\lambda_{S_t}}\left(\lambda_{S_t}^2+ \lambda_{P_t}^2+\frac{3}{2} \lambda_{T_t}^2\right)\nonumber   
   \\&
\qquad\qquad\quad +6 (\lambda_{S}+ \lambda_{T})+2\lambda_{P}
   -\tau N_f T_F \lambda_{S_t}\nonumber   
   \\&
\qquad\qquad\quad    +\left(\frac{3g^2\kappa^2}{2}+6\lambda_{T_t}\right)\frac{\lambda_{T}}{\lambda_{S_t}}+
   4 \lambda_{P}\frac{\lambda_{P_t}}{\lambda_{S_t}}
\bigg\}\frac{1}{\tilde{\epsilon}},\nonumber
\\
Z_{Pq^2}=&1-\frac{1}{(4\pi)^2}\bigg\{C_F\left(C_A-2C_F\right)\frac{1}{\lambda_{P}}\left[\frac{3g^2\kappa^2}{4}\lambda_{T_t}+2 \lambda_{P_t} \lambda_{S_t}+\frac{3}{2}
   \lambda_{T_t}^2\right]\label{ZPq^2}\\&
\qquad\qquad\quad+2C_F\left(2g^2\xi+\frac{3g^2\kappa^2}{4}+\lambda_{S_t}+\lambda_{P_t}+3 \lambda_{T_t}\right)\nonumber
\\&
\qquad\qquad\quad+ \lambda_{P} (2-\tau N_f N_c )+6 (\lambda_{S}+\lambda_{T})
+3 \frac{\lambda_{T}^2}{\lambda_{P}}
 \bigg\}\frac{1}{\tilde{\epsilon}},\nonumber
\end{align}
\begin{align} 
Z_{P_tq^2}=&1-\frac{1}{(4\pi)^2}\bigg\{\left(4C_F-\frac{3}{2}C_A\right)\frac{1}{\lambda_{P_t}}\left[\frac{3g^2\kappa^2}{4}\lambda_{T_t}+2\lambda_{S_t}\lambda_{P_t}+\frac{3}{2} \lambda_{T_t}^2\right]\label{ZPtq^2}\\&
\qquad\qquad\quad+2C_Fg^2\xi+\left(2C_F-C_A\right)\left(\frac{3g^2\kappa^2}{4}+\lambda_{S_t}+\lambda_{P_t} +3\lambda_{T_t}\right)\nonumber
   \\&
\qquad\qquad\quad+6 (\lambda_{S}+\lambda_{T})+2\lambda_{P}
   -\tau N_f T_F \lambda_{P_t}\nonumber
   \\&
\qquad\qquad\quad
   +\left(\frac{3g^2\kappa^2}{2}+6\lambda_{T_t}\right)\frac{\lambda_{T}}{\lambda_{P_t}}+
   4 \lambda_{P}\frac{\lambda_{S_t}}{\lambda_{P_t}}
\bigg\}\frac{1}{\tilde{\epsilon}},\nonumber
\\
Z_{Tq^2}=&1-\frac{1}{(4\pi)^2}\Bigg\{C_F\left(C_A-2C_F\right)\frac{1}{\lambda_{T}}\left(\frac{g^2\kappa^2}{2}+ 2\lambda_{T_t} \right)\left(\lambda_{S_t}+\lambda_{P_t}\right)\label{ZTq^2}\\&
\qquad\qquad\quad+  2C_F\left(2g^2\xi-\frac{g^2\kappa^2}{4}+\lambda_{P_t}
   +\lambda_{S_t}- \lambda_{T_t}\right)\nonumber
\\&
\qquad\qquad\quad+ 6 ( \lambda_{S}+ \lambda_{P})-2\lambda_{T}\left(1+
   \frac{\tau}{4}N_f N_c\right)\Bigg\}\frac{1}{\tilde{\epsilon}}+\lambda_{T}+\ft,\nonumber
\\
Z_{T_tq^2}=&1-\frac{1}{(4\pi)^2}\bigg\{C_A g^2\left[\frac{g^2\kappa^2}{4\lambda_{T_t}}\left(1-\frac{\kappa}{2}\right)^2-\frac{3\kappa^2}{4\lambda_{T_t}}\left(\lambda_{S_t}+\lambda_{P_t}-\lambda_{T_t}\right)\right]\label{ZTtq^2}\\&
\qquad\qquad\quad+C_Fg^2\left[2\xi+\frac{\kappa^2}{2\lambda_{T_t}}\left(4\lambda_{S_t}+4\lambda_{P_t}-\lambda_{T_t}\right)\right]+\frac{g^2\kappa^2}{\lambda_{T_t}}\left(\lambda_{S}+\lambda_{P}\right)\nonumber	\\
&\qquad\qquad\quad+2C_A\left(\lambda_{T_t}-2\lambda_{S_t}-2\lambda_{P_t}\right)+ 2C_F\left(5\lambda_{S_t}+5\lambda_{P_t}-\lambda_{T_t}\right)\nonumber\\
&\qquad\qquad\quad+6\left(\lambda_{S}+\lambda_{P}\right)-2\lambda_{T}-\frac{\tau}{2}N_fT_F\lambda_{T_t}+4 \frac{\lambda_{T}}{\lambda_{T_t}}(\lambda_{S_t}+\lambda_{P_t} ) 
\bigg\}\frac{1}{\tilde{\epsilon}}.\nonumber
\end{align}

\section{Beta functions}\label{bfss}
Form the renormalization constants obtained in eqs. (\ref{ZA}), (\ref{Zq}), (\ref{Zmq^2}), (\ref{Zc}), (\ref{ZAq^2}), (\ref{ZkAq^2}), (\ref{ZA^2q^2}), (\ref{ZkA^2q^2}), (\ref{ZAc^2}), (\ref{ZA^3}), (\ref{ZA^4}), (\ref{ZSq^2}), (\ref{ZStq^2}), (\ref{ZPq^2}), (\ref{ZPtq^2}), (\ref{ZTq^2}), (\ref{ZTtq^2})
and their definition in eq. (\ref{count1}), it can be shown that the following Slavnov-Taylor identity hold at one loop:
\begin{equation}
\begin{split}
\frac{Z_{Aq^2}}{Z_q}=\frac{Z_{A^2q^2}}{Z_{Aq^2}}=\frac{Z_{\kappa A^2q^2}}{Z_{\kappa Aq^2}}=\frac{Z_{Ac^2}}{Z_c}=\frac{Z_{A^3}}{Z_A}=\frac{Z_{A^3}}{Z_A}=\frac{Z_{A^4}}{Z_{A^3}}=1-\frac{g^2}{(4\pi)^2}C_A\left(\frac{3+\xi}{4}\right)\frac{1}{\tilde{\epsilon}}.
\end{split}
\end{equation}

The beta functions $\beta_\eta\equiv\mu\frac{\partial\eta}{\partial\mu}$ and anomalous dimensions $\gamma_m\equiv\frac{\mu}{m}\frac{\partial m}{\partial\mu}$  in the $\epsilon\to 0$ limit can be extracted from the obtained renormalization constants as
\begin{align}
\beta_g =&-\frac{g^3 }{(4\pi)^2}\left[\frac{11}{3} C_A-\tau N_f T_F\left(\frac{3 \kappa ^2-4}{12}\right)  \right],\label{bg}\\
\beta_{\kappa}=&\frac{ \kappa }{(4 \pi )^2}\bigg\{\frac{g^2}{4} \left[2C_F( \kappa ^2-4)-C_A (\kappa -2) (\kappa +6) \right]-\tau N_f T_F\lambda_{T_t} \label{bk}\\
&\qquad\qquad+(C_A-2 C_F) (\lambda_{S_t}+\lambda_{P_t}-\lambda_{T_t})-2 (\lambda_{S}+\lambda_{P}-\lambda_{T}) \bigg\},\nonumber
\end{align}
\begin{align} 
\beta_{\lambda_S}=&-\frac{1}{(4\pi)^2}\bigg\{\frac{3g^2}{8} \left[g^2 \left(\kappa ^2-4\right)^2+4\kappa^2 \lambda_{T_t} \right]C_F(C_A-2 C_F)\label{bS}\\
&\qquad\qquad+C_F( C_A-2 C_F)  \left(2\lambda_{S_t}^2+2\lambda_{P_t}^2+3\lambda_{T_t}^2\right)\nonumber\\
&\qquad\qquad+C_F\lambda_{S}\left[3g^2 \left(\kappa ^2+4\right) +4(\lambda_{P_t}+\lambda_{S_t}+3 \lambda_{T_t})\right]\nonumber\\
&\qquad\qquad+2 \left[2
   \lambda_{P}^2+2 \lambda_{P} \lambda_{S}+\lambda_{S}^2 (4-\tau N_c N_F )+6 \lambda_{S} \lambda_{T}+3 \lambda_{T}^2\right]\bigg\},\nonumber\\
\beta_{\lambda_{S_t}}=&-\frac{1}{(4\pi)^2}\bigg\{\frac{3g^2}{8} \left[g^2 \left(\kappa ^2-4\right)^2+4\kappa^2 \lambda_{T_t} +32 \lambda_{S_t}\right]\left(4 C_F-\frac{3}{2} C_A\right)\\
&\qquad\qquad+ \left(4 C_F-\frac{3}{2}C_A\right) \left(2\lambda_{S_t}^2+2\lambda_{P_t}^2+3 \lambda_{T_t}^2 \right)+ 3g^2 \kappa ^2 \lambda_{T}\nonumber\\
&\qquad\qquad+
   \lambda_{S_t} \left( C_F-\frac{1}{2}C_A\right)\left[3g^2\left(\kappa ^2-12\right)+ 4(\lambda_{P_t}+\lambda_{S_t}+3 \lambda_{T_t})\right]\nonumber\\
&\qquad\qquad+2\left[2
   \lambda_{P}(2 \lambda_{P_t}+ \lambda_{S_t})+6\lambda_{S} \lambda_{S_t}-  \tau N_f T_F\lambda_{S_t}^2+6\lambda_{T}( \lambda_{S_t} +
    \lambda_{T_t})\right]\bigg\},\nonumber
\\  
\beta_{\lambda_P}=&-\frac{1}{(4\pi)^2}\bigg\{C_F\left(C_A-2C_F\right)\left(\frac{3 g^2\kappa ^2}{2}  \lambda_{T_t}+4 \lambda_{P_t} \lambda_{S_t}+3 \lambda_{T_t}^2\right)\\
&\qquad\qquad+C_F\lambda_{P}\left[3 g^2
   \left(\kappa ^2+4\right) +4 (\lambda_{P_t}+\lambda_{S_t}+3 \lambda_{T_t})\right]\nonumber\\
& \qquad\qquad+2 \left[\lambda_{P}^2 (2- \tau N_c N_F )+6 \lambda_{P}
   (\lambda_{S}+\lambda_{T})+3 \lambda_{T}^2\right]\bigg\},\nonumber\\
\beta_{\lambda_{P_t}}=&-\frac{1}{(4\pi)^2}\bigg\{ \left(4 C_F -\frac{3}{2} C_A\right) \left[\frac{3 g^2}{2} \left(8 \lambda_{P_t}+\kappa ^2 \lambda_{T_t}\right)+4 \lambda_{P_t} \lambda_{S_t}+3 \lambda_{T_t}^2\right]\\
&\qquad\qquad+\left( C_F -\frac{1}{2} C_A\right)
   \lambda_{P_t} \left[3 g^2 \left(\kappa ^2-12\right)+4 (\lambda_{P_t}+\lambda_{S_t}+3 \lambda_{T_t})\right]\nonumber\\
&\qquad\qquad+2\left[2 \lambda_{P}(\lambda_{P_t}+2 \lambda_{S_t})- \tau N_f T_F \lambda_{P_t}^2 +6 \lambda_{P_t} \lambda_{S}+6\lambda_{T}( \lambda_{P_t} +\lambda_{T_t})\right]\nonumber\\
&\qquad\qquad +3 g^2 \kappa ^2 \lambda_{T}
\bigg\},\nonumber\\
\beta_{\lambda_{T}}=&-\frac{1}{(4\pi)^2}\bigg\{C_F(C_A -2 C_F) (\lambda_{P_t}+\lambda_{S_t})
   \left(g^2 \kappa ^2+4 \lambda_{T_t}\right)\\
&\qquad\qquad+C_F \lambda_{T}\left[g^2(12-\kappa ^2) +4(\lambda_{P_t} +\lambda_{S_t} -\lambda_{T_t})\right]\nonumber\\
&\qquad\qquad+\lambda_{T} \left[12 \lambda_{P}+12 \lambda_{S}-\lambda_{T} ( \tau N_c N_f +4)\right]\bigg\},\nonumber\\
\beta_{\lambda_{T_t}}=&-\frac{1}{(4\pi)^2}\bigg\{ C_A g^2 \left[\frac{1}{8} g^2 (\kappa -2)^2 \kappa ^2-\frac{3 \kappa ^2}{2} (\lambda_{P_t}+\lambda_{S_t}-\lambda_{T_t})\right]\label{bTt}\\
&\qquad\qquad+4 C_A \lambda_{T_t} \left[\lambda_{T_t}-2 (\lambda_{P_t}+\lambda_{S_t})\right]+C_F g^2 \kappa ^2 \left[4 (\lambda_{P_t}+\lambda_{S_t})-\lambda_{T_t}\right]\nonumber\\
&\qquad\qquad+4 C_F \lambda_{T_t} \left[3 g^2+5 (\lambda_{P_t}+\lambda_{S_t})-\lambda_{T_t}\right]+2 (\lambda_{P}+\lambda_{S}) \left(g^2 \kappa ^2+6 \lambda_{T_t}\right)\nonumber\\
&\qquad\qquad+8 \lambda_{T} (\lambda_{P_t}+\lambda_{S_t})-4 \lambda_{T} \lambda_{T_t}-\lambda_{T_t}^2
   \tau N_f T_F\bigg\},\nonumber\\
\gamma_{m}=&-\frac{1}{(4\pi)^2}\bigg\{ C_F\left[\frac{3 g^2}{4} \left(\kappa ^2+4\right)+ (\lambda_{P_t}+\lambda_{S_t}+3 \lambda_{T_t})\right]\\
&\qquad\qquad +(\lambda_{P}-
   \tau N_c N_f \lambda_{S} +\lambda_{S}+3 \lambda_{T})\bigg\}.\nonumber
\end{align}

Notice that there is a fixed point for eqs. (\ref{bk})--(\ref{bTt}) at $\kappa=2$ with all self-interaction couplings set to zero. In this fixed point, the theory has a simple connection to Dirac QCD, which amounts to take formally $\tau\to 2$ (See \cite{VNA} for an analogous discussion in QED). This is the only choice available for pure second order QCD (vanishing self-interaction couplings).

\section{Summary and conclusions}\label{sectSumm}
In this work, the one-loop level renormalization of QCD in the second order Poincar\'e projector formalism was performed. The description of an arbitrary chromomagnetic factor is natural in this framework and fermion self-interactions can be incorporated without spoiling renormalizability. As expected, the conclusions of this analysis are analogous to those of \cite{VNA}, and can be summarized as follows:
$i)$ Pure second order QCD is renormalizable only for $\kappa=2$. 
$ii)$ Fermion self-interactions are also renormalizable at one-loop and make possible the renormalization of an arbitrary chromomagnetic factor in the second order formalism. Finally it is important to remark that the last conclusion requires a deeper analysis due to the $\gamma^5$ issues of the dimensional regularization method.

\acknowledgments{This work was supported by CONACyT under project CB-2011-167425. The author acknowledges Mauro Napsuciale and Alfredo Aranda for their helpful comments and critical reading of the manuscript.}

\end{document}